\documentclass[journal]{IEEEtran}

\usepackage{stfloats}
\usepackage{hhline}
\usepackage[numbers]{natbib}
\usepackage[utf8]{inputenc}
\usepackage{siunitx}
\usepackage{graphicx}
\usepackage{makecell}
\usepackage{comment}
\usepackage[dvipsnames]{xcolor}
\usepackage{adjustbox}
\usepackage{hyperref}
\usepackage{subcaption}
\usepackage{caption}
\usepackage{array}
\usepackage{algorithm,algorithmic}

\begin{document}
%
\title{Studying the Robustness of Anti-adversarial Federated Learning Models Detecting Cyberattacks in IoT Spectrum Sensors}

\author{Pedro Miguel S\'anchez S\'anchez$^{1}$, Alberto Huertas Celdr\'an$^{*2}$, Timo Schenk$^{2}$, Adrian Lars Benjamin Iten$^{2}$,\\ G\'er\^ome Bovet$^{3}$, Gregorio Mart\'inez P\'erez$^{1}$, and Burkhard Stiller$^{2}$

\thanks{$^{*}$Corresponding author.}

\thanks{$^{1}$Pedro Miguel S\'anchez S\'anchez and Gregorio Mart\'inez P\'erez are with the Department of Information and Communications Engineering, University of Murcia, 30100 Murcia, Spain {\tt\small (pedromiguel.sanchez@um.es; gregorio@um.es)}.}%
\thanks{$^{2}$Alberto Huertas Celdr\'an, Timo Schenk, Adrian Lars Benjamin Iten, and Burkhard Stiller are with the Communication Systems Group (CSG) at the Department of Informatics (IfI), University of Zurich UZH, 8050 Zürich, Switzerland {\tt\small (e-mail: huertas@ifi.uzh.ch; timo.schenk@uzh.ch; adrianlarsbenjamin.iten@uzh.ch; stiller@ifi.uzh.ch}).}
\thanks{$^{3}$G\'{e}r\^{o}me Bovet is with the Cyber-Defence Campus within armasuisse Science \& Technology, 3602 Thun, Switzerland {\tt\small (gerome.bovet@armasuisse.ch)}.}}


\markboth{IEEE Transactions on Dependable and Secure Computing}%
{Shell \MakeLowercase{\textit{et al.}}: Bare Demo of IEEEtran.cls for IEEE Journals}

\maketitle

\begin{abstract}
Device fingerprinting combined with Machine and Deep Learning (ML/DL) report promising performance when detecting cyberattacks targeting data managed by resource-constrained spectrum sensors. However, the amount of data needed to train models and the privacy concerns of such scenarios limit the applicability of centralized ML/DL-based approaches. Federated learning (FL) addresses these limitations by creating federated and privacy-preserving models. However, FL is vulnerable to malicious participants, and the impact of adversarial attacks on federated models detecting spectrum sensing data falsification (SSDF) attacks on spectrum sensors has not been studied. To address this challenge, the first contribution of this work is the creation of a novel dataset suitable for FL and modeling the behavior (usage of CPU, memory, or file system, among others) of resource-constrained spectrum sensors affected by different SSDF attacks. The second contribution is a pool of experiments analyzing and comparing the robustness of federated models according to \textit{i)} three families of spectrum sensors, \textit{ii)} eight SSDF attacks, \textit{iii)} four scenarios dealing with unsupervised (anomaly detection) and supervised (binary classification) federated models, \textit{iv)} up to 33\% of malicious participants implementing data and model poisoning attacks, and \textit{v)} four aggregation functions acting as anti-adversarial mechanisms to increase the models robustness. 

\end{abstract}


\IEEEpeerreviewmaketitle

\section{Introduction}
\label{sec:intro}

In crowdsensing, large groups of individuals collaborate in a crowdsourcing fashion, typically leveraging devices as sensors~\cite{liu2018survey}. Employing resource-constrained spectrum sensors (Raspberry Pis equipped with software-defined radio kits), the ElectroSense initiative marks an exemplary network for crowdsensing in particular~\cite{rajendran2017electrosense}. However, the rapid growth of spectrum sensors also has accelerated the emergence of new and specialized cyberattacks, called spectrum sensing data falsification (SSDF) attacks~\cite{yadav:2020:ssdf}. In such a context, the privacy and integrity of sensors measurements are at risk.

In order to detect SSDF attacks affecting resource-constrained sensors, signature-based approaches present the limitation of not being effective against new attacks that have not been observed during the signature creation stage (zero-day attacks). To overcome this limitation, dynamic anomaly detectors considering fingerprinting are gaining relevance. This approach monitors device activities such as the usage of CPU, memory, network interfaces, or file system when there is no infection, and in a second stage, detects deviations produced by SSDF attacks~\cite{sanchez2021survey}. The detection phase can be implemented using different techniques. One of the most lightweight in terms of resource consumption is based on rules, but the creation of precise rules requires expert knowledge and a relevant amount of time in complex crowdsensing scenarios \cite{hamza2018combining}. Knowledge-based solutions have also been proposed in the literature, but they do not scale well with many sensors, requiring a lot of time to model and detect attacks \cite{khraisat2019survey}. Finally, machine and deep learning (ML/DL) techniques are gaining enormous relevance due to their detection performance, time, and relative simplicity \cite{aldweesh2020deep}.

Despite the benefits of anomaly detectors combining device fingerprinting and ML/DL, they present some characteristics limiting their applicability in crowdsensing scenarios where data belongs to different sensors, and it cannot be shared due to privacy restrictions. Thus, federated learning (FL) becomes increasingly relevant~\cite{yang2019federated}. Compared to centralized approaches, FL aims to train a federated model collaboratively but in a decentralized and privacy-preserving fashion. Each participant of the federation trains a model with its own data and shares the model parameters to create the federated model. However, FL also presents security concerns, being adversarial attacks launched by malicious participants one of the most important ones. In this context, the literature has proposed several data and model falsification attacks consisting of poisoning data, labels, or weights during training to exchange fake model parameters with the entity (or entities) creating the federated model \cite{rodriguezbarroso2022survey}. To overcome this problem, different countermeasures have been proposed, such as the usage of secure aggregation functions \cite{pillutla2019robust}. However, due to the novelty of the field, the combination of FL and behavioral fingerprinting for detecting SSDF attacks on spectrum sensor devices poses several open challenges. 

First, there is an evident lack of FL-oriented datasets modeling fingerprints of resource-constrained devices belonging to real platforms~\cite{rey2022federated}. Second, there is no work measuring the performance of FL models using device fingerprinting to detect SSDF attacks affecting spectrum sensors and comparing its detection performance with existing traditional ML/DL-based solutions. Last but not least, there is no work studying the robustness of FL-based solutions oriented to detect SSDF attacks in spectrum sensors and equipped with different anti-adversarial mechanisms to mitigate the impact of heterogeneous data and model poisoning attacks.

To improve the previous challenges, this paper presents the following contributions:

\begin{enumerate}
  \item The creation of a novel device behavioral fingerprinting dataset suitable for FL scenarios (publicly available in~\cite{dataset}). This dataset contains the normal and under-attack behavior of four ElectroSense spectrum sensors, which are implemented with three different families of Raspberry Pis connected to software-defined radio kits. About 75 internal events belonging to the usage of CPU, memory, network interface, file systems, and other relevant dimensions are monitored in each sensor for two different versions of normal behavior as well as eight SSDF attacks.
  
  \item The usage of the dataset to conduct a pool of experiments evaluating and comparing the performance of \textit{(i)} DL models under a horizontal FL scheme, and \textit{(ii)} traditional DL approaches where a centralized aggregation neglects privacy. This evaluation comprises the definition of four federated scenarios dealing with anomaly detection (using Autoencoder), binary classification (with multilayer perceptron), and different participants. 
  
  \item The study of the federated models robustness in two of the previous federated scenarios under different conditions. The conditions vary in terms of \textit{(i)} anti-adversarial aggregation mechanisms (two variants of \textit{trimmed mean}, and \textit{coordinate-wise median}), \textit{(ii)} an increasing number of malicious participants (from 8 to 33\%), and \textit{(iii)} heterogeneous data and model poisoning attacks affecting both supervised and unsupervised FL models.

\end{enumerate}

The remainder of this article is organized as follows. Section~\ref{sec:related} reviews solutions combining fingerprinting and ML/DL/FL approaches able to detect cyberattacks affecting IoT. While Section~\ref{sec:dataset} provides the details of the FL-oriented dataset created in this work, Section~\ref{sec:federated} evaluates and compares the performance of different FL and traditional DL models trained and evaluated in heterogeneous conditions and scenarios. Section~\ref{sec:attacks} analyzes the robustness of FL models affected by different adversarial attacks. Finally, Section~\ref{sec:conclusions} draws conclusions and next steps.

\section{Related Work}
\label{sec:related}

This section reviews related work considering behavioral fingerprinting and ML/DL approaches, both centralized and federated, to detect cyberattacks affecting IoT.

In~\cite{sanchez2021survey} a broad survey of device fingerprinting reviews a good number of works detecting cybersecurity issues in IoT devices. One of the main conclusions of this survey is that there is a current trend towards combining device fingerprinting and ML/DL/FL techniques to detect cybersecurity attacks. In this context, the work most related to the paper at hand in terms of attacks, devices, and behavioral events is proposed in~\cite{huertas2022cyberspec}. The authors combine unsupervised ML/DL techniques and the usage of device resources (such as CPU, memory, file system, or the network interface, among others) to detect anomalies produced by seven SSDF attacks affecting different Raspberries Pi acting as ElectroSense sensors. A pool of experiments reports 80-100\% TPR when detecting five of the seven SSDF attacks. In~\cite{blaise2020botfp}, the authors look at frequency distributions of protocol attributes and run clustering algorithms to capture particularities of botnet behaviors. They report 97-100\% accuracy, and as in most works, networking features are leveraged as the behavioral source. The authors of~\cite{kumar2019edima} use ML techniques combined with network packets to detect heterogeneous malware in IoT devices. They achieve 95\% accuracy on their test sets. Comparing the previous three works and the paper at hand, the main difference is that the proposed ML/DL models are created in a centralized manner, which means that the privacy of the data used to train the models has not been guaranteed, one of the main contributions of this work.

Dealing with solutions that use FL to detect malware affecting IoT devices in privacy-preserving scenarios. The authors of~\cite{taheri2020fed} propose a solution for industrial IoT that analyzes Android application samples and behavioral data in an industrial context. They report 97-100\% accuracy when detecting different malware samples. \cite{preuveneers2018chained} presents a different use case for FL in the field of intrusion detection with up to 97\% accuracy. This work studies adversarial implications in FL and employs blockchain technology as an alternative to mitigate them. Therefore, this work focuses more on the accountability of participants instead of reducing the impact of the attacks. The main difference between the previous approaches and the one proposed in this work is that they do not consider behavioral fingerprinting, do not consider the problem of malicious participants, and do not evaluate the robustness of their models against adversarial attacks.

Most related to this work, several works combine FL and device fingerprinting to detect cyberattacks. \cite{hsu2020privacy} presents an FL system to detect malware in Android. The authors train Support Vector Machine classifiers in a federated scenario using device features such as application programming interface (API) calls and permission configuration to obtain 94-96\% F1-score. While API calls correspond to device behavioral source fingerprints, the work at hand analyzes behavioral data sources on a much lower level. In~\cite{nguyen2019diot}, a federated anomaly detection system is proposed for the IoT. It leverages device-type profiles of the communications to detect malware with 96\% accuracy. In contrast to the work at hand, the previous solution is not effective against malware affecting data availability, integrity or confidentiality. The authors of~\cite{rey2022federated} leverage the N-BaIoT dataset to train and evaluate FL models. They achieve good accuracy in federated scenarios dealing with network traffic events. In addition to that, adversarial impacts are measured for selected attacks, and different mechanisms of robust aggregation are evaluated. While the paper at hand considers adversarial attacks against federations, it applies and analyzes these concepts in a scenario with very different characteristics to the one used in~\cite{rey2022federated}. In this sense, N-BaIoT contains network traffic from nine different IoT devices such as webcams and smart doorbells. However, N-BaIoT does not consider spectrum sensors such as Raspberry Pis equipped with software-defined radio kits (as this work does), and does not model device behavioral fingerprinting events.

\begin{table*}[bp]
\caption{Comparison of Related Work}
\centering
    \resizebox{\textwidth}{!}{\begin{tabular}{c|c|c|c|c|c|c|c}
Source                                         & Device Types           & Attack Type     & Data/Fingerprints      & ML Approach & Prediction                             & Privacy & Robustness            \\ \hline\hline
\cite{huertas2022cyberspec}       & Raspberry Pis            & SSDF        & Usage of Resources    & ML/DL      & Anomaly Detection   & no     & no           \\
\cite{blaise2020botfp}        & Multiple               & Botnets         & Communication-based    & ML          & Classification (Distances to clusters) & no      & no                    \\
\cite{kumar2019edima}         & IoT devices            & Botnets         & Communication-based    & ML          & Classification                         & no      & no                    \\
\cite{taheri2020fed}          & Industrial IoT devices & Android Malware & App Information & FL, DL      & Classification                         & yes     & yes, GAN-based        \\
\cite{preuveneers2018chained} & Computers/Machines     & multiple        & Communication-based    & FL, DL      & Anomaly Detection (Autoencoder)        & yes     & yes, blockchain-based \\
\cite{hsu2020privacy}         & Mobile (Android)       & Android Malware & App Information        & FL, ML      & Classification (SVM)                   & yes     & no                    \\
\cite{nguyen2019diot}         & IoT devices            & IoT Malware     & Communication-based    & FL, ML      & Anomaly Detection                      & yes     & no                    \\
\cite{rey2022federated}       & IoT devices            & Botnets         & Communication-based    & FL, DL      & Anomaly Detection and Classification   & yes     & yes, aggregation                   \\ 
ours       & Raspberry Pis            & SSDF         & Usage of Resources    & FL, DL      & Anomaly Detection and Classification   & yes     & yes, aggregation                   \\\hline
\end{tabular}}
\label{tab:related}
\end{table*}

As can be seen in \tablename~\ref{tab:related}, none of the related work studies the detection performance and robustness of federated models detecting SSDF attacks. Only~\cite{huertas2022cyberspec} covers the same attacks and devices considered in this work, but from a traditional ML/DL perspective and without considering privacy-preserving scenarios. Moreover, \tablename~\ref{tab:related} shows that device behavioral fingerprints have not been used for the use case of federated malware detection. In conclusion, this literature review demonstrates the lack of works and datasets combining behavioral fingerprints and FL to detect cyberattacks in IoT devices similar to those used in crowdsensing platforms. Furthermore, to the best of our knowledge, there is no work studying the impact of adversaries on the robustness of the previous federated models.


\section{Dataset Creation}
\label{sec:dataset}

This section describes the novel device fingerprinting dataset created for federated scenarios. In particular, it presents \textit{(i)} the crowdsensing platform and the spectrum sensors used to create the dataset, \textit{(ii)} the SSDF attacks affecting the deployed ElectroSense sensors, \textit{(iii)} the events selected to create the fingerprints, and \textit{(iv)} the exploration of the dataset content.

\subsection{ElectroSense Sensors \& SSDF Attacks}

ElectroSense is a real and collaborative crowdsensing platform that pursues the goal of monitoring the electromagnetic space \cite{electrosense}. ElectroSense is composed of a multitude of spectrum sensors built from cheap commodity hardware like Raspberry Pis equipped with software-defined radio scanners and antennas. Each sensor monitors the different bands and segments of the radio frequency spectrum within its location. Sensed spectrum data is periodically sent to a backend platform in charge of storing, pre-processing, and analyzing the data to provide services. These services range from spectrum occupancy monitoring to transmission decoding. In this scenario, four physical spectrum sensors have been deployed in two locations. \tablename~\ref{tab:setup-devices} summarizes the devices identifiers, hardware characteristics, and locations.

\begin{table}[]
    \caption{Details of the Devices Making up the Scenario}
    \centering
    \begin{tabular}{c|c|c|c}
        Device ID & Type/Model & RAM & Location  \\ \hline
        \hline
        RPi3 & 3 Model B+ & 1GB & Zurich  \\ 
        RPi4\_1 & 4 Model B & 2GB & Zug   \\ 
        RPi4\_2 & 4 Model B & 2GB & Zug  \\ 
        RPi4\_3 & 4 Model B & 4GB & Zurich  \\
        \hline
    \end{tabular}
    \label{tab:setup-devices}
\end{table}

For each ElectroSense sensor, two versions of the official and publicly available software are used. The first version is the current sensing application, installed by default in the sensor. The second version of the ElectroSense sensor software is an old one, available on the official ElectroSense GitHub \cite{essensor}. Additionally, eight different SSDF attacks are considered to infect the four sensors. These SSDF attacks are executed after modifying the ElectroSense sensor source code and compiling a new version of the executable. The main goal of these SSDF attacks is to manipulate the data of particular spectrum segments monitored by the sensors (in different ways) and send poisoned spectrum data to the ElectroSense backend platform. Despite the differences in terms of attacks impacts, all attacks affect the same number of spectrum segments (20 MHz). \tablename~\ref{tab:attacks} summarizes the main aspects of the behaviors considered during the creation of the dataset. More details about the implementation and functionality of the SSDF attacks can be found in \cite{huertas2022cyberspec}.

\begin{table}[ht]
    \scriptsize
    \caption{Behaviors Monitored During the Dataset Creation}
    \centering
    \begin{tabular}{c|>{\raggedright\arraybackslash}m{5.8cm}|c}
        Behavior & \makecell[c]{Description} & Time \\
        \hline
        \hline
        Normalv1 & Current ElectroSense application sensing the spectrum & 5 days \\
        \hline
        Normalv2 & Old ElectroSense application sensing the spectrum & 5 days \\
        \hline
        Delay & Sense different outdated spectrum data of affected segments & 4 hours \\
        \hline
        Confusion & Swap the spectrum data between affected segments & 4 hours \\
        \hline
        Freeze & Sense the same outdated spectrum data in affected segments & 4 hours \\
        \hline
        Hop & Add random noise to random parts of affected segments & 4 hours \\
        \hline
        Mimic & Copy the spectrum data of one segment into another segment & 4 hours \\
        \hline
        Noise & Add random noise to the spectrum data of affected segments & 4 hours \\
        \hline
        Repeat & Replicate the same spectrum data in all affected segments & 4 hours \\
        \hline
        Spoof & Copy the spectrum data of one segment into another segment and add random noise & 4 hours \\
        \hline
    \end{tabular}
    \label{tab:attacks}
\end{table}

\begin{figure}[htpb!]
    \centering
    \includegraphics[width=0.9\columnwidth]{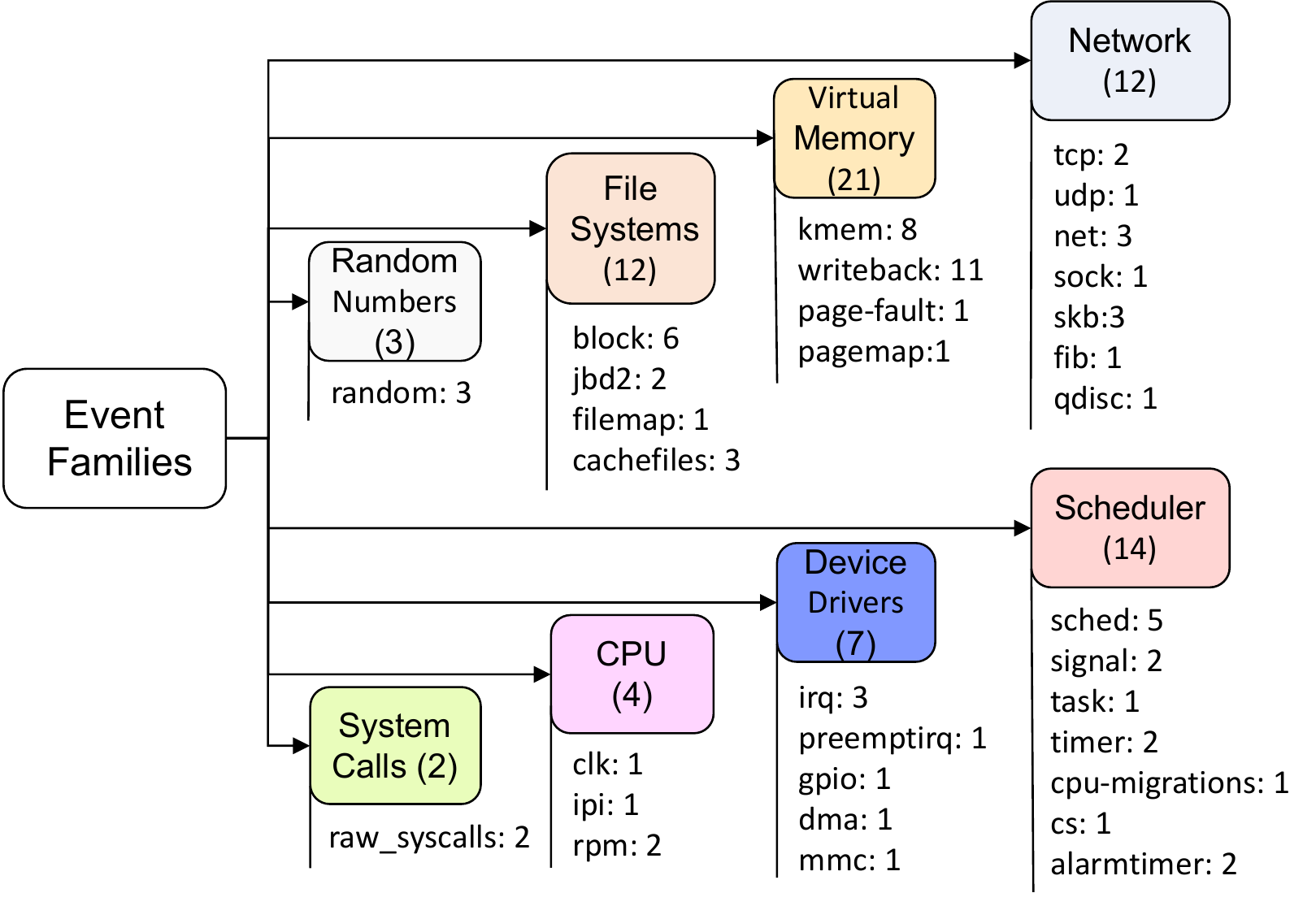}
    \caption{Device Fingerprinting Events of the FL-oriented Dataset}
    \label{fig:events}
\end{figure}

The previous behaviors are sequentially executed in the devices of \tablename~\ref{tab:setup-devices} for five days (the normal behaviors) and four hours (the attacks), as indicated in \tablename~\ref{tab:attacks}. To create the fingerprinting dataset, 75 internal events of each device have been monitored in time windows of 50 s using the \textit{perf} Linux command. These events belong to the following device data sources: CPU, virtual memory, network, file system, scheduler, device drivers, and random numbers. The number of events, per event type and family, contained in the datasets can be seen in \figurename~\ref{fig:events}. As a summary, the dataset includes four ElectroSense sensors, ten behaviors (two normal and eight SSDF attacks), 75 events belonging to eight event families, and a total of 73396 samples (approximately 60000 samples of normal and 13396 of malicious behavior). The dataset is publicly available in~\cite{dataset}.

\subsection{Data Exploration}
\label{sub:data_exploration}

This section explores the created dataset to find data patterns and determine the suitability of ML/DL/FL techniques to detect SSDF attacks. This analysis also aims to determine if the data contained in the dataset is independent and identically distributed (IID) or non-IID. For that, three types of studies are performed. The first analyzes the evolution of data over time. The second focuses on the distributions of data belonging to different devices. Finally, the third explores data distributions according to various SSDF attacks.

\begin{figure}[htpb!]
    \centering
    \includegraphics[width=\linewidth]{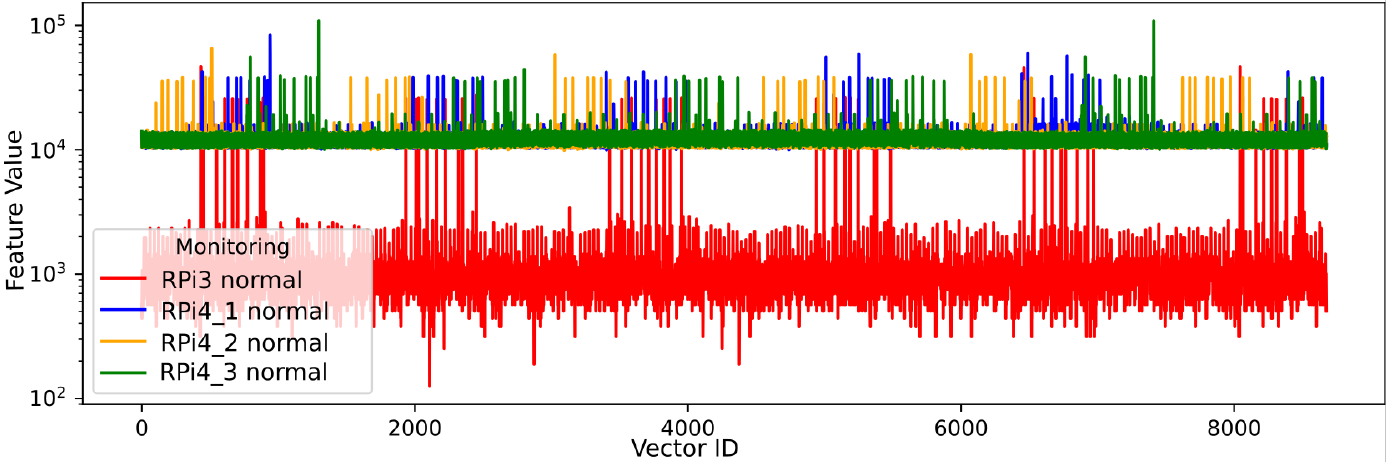}
    \caption{\textit{kmen:mm\_page\_pcpu\_drain} event for Normal Behavior in all Devices}
    \label{fig:normal-hist}
\end{figure}

The variation of behavioral data over time is essential to determine the stability of fingerprints, and the suitability of ML/DL/FL approaches to detect normal behavior and SSDF attacks. In this context and as an example, \figurename~\ref{fig:normal-hist} shows the values of the \textit{kmen:mm\_page\_pcpu\_drain} event belonging to the \textit{Virtual Memory} family across the time and for each device. As can be seen, the values are periodic, with some repetitive peaks. Exploring more in detail \figurename~\ref{fig:normal-hist}, it is also interesting to see the different distribution for RPi3 (in red) and RPi4s (in blue, orange, and green), indicating that behavioral data of similar devices is IID, and for different devices is non-IID. In particular, the range of values of the \textit{kmen:mm\_page\_pcpu\_drain} event for the RPi3 is different from the range for the RPi4 devices. These characteristics are also visible in the majority of events, but they are not included due to room constraints.

\begin{figure}[htpb!]
    \centering
    \includegraphics[width=\linewidth]{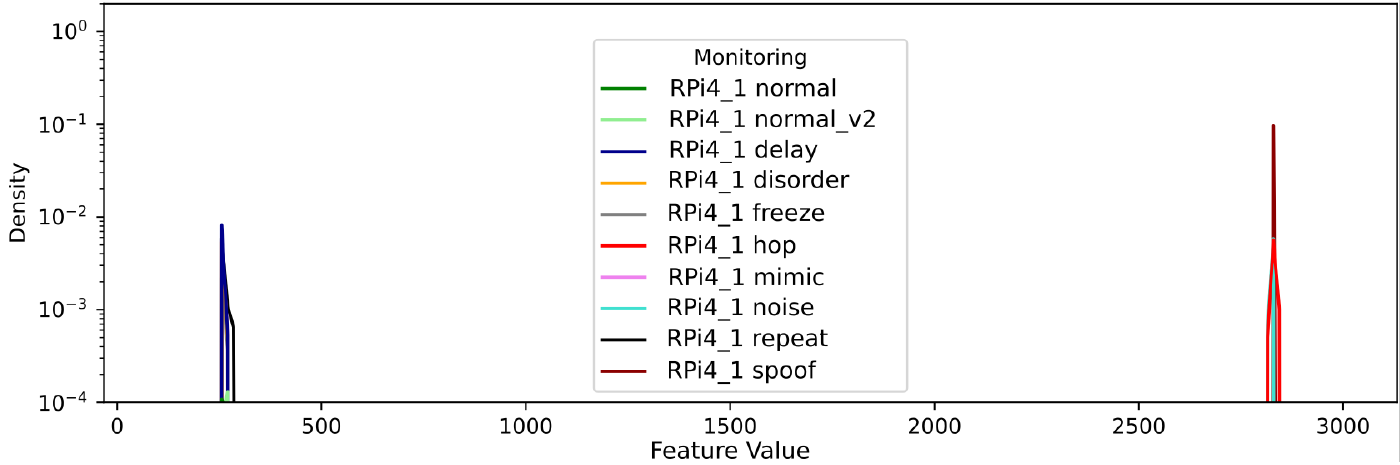}
    \caption{\textit{urandon\_read} Event of all Behaviors on RPi4\_1}
    \label{fig:attacks-randomness}
\end{figure}

\begin{figure}[htpb!]
    \centering
    \includegraphics[width=\linewidth]{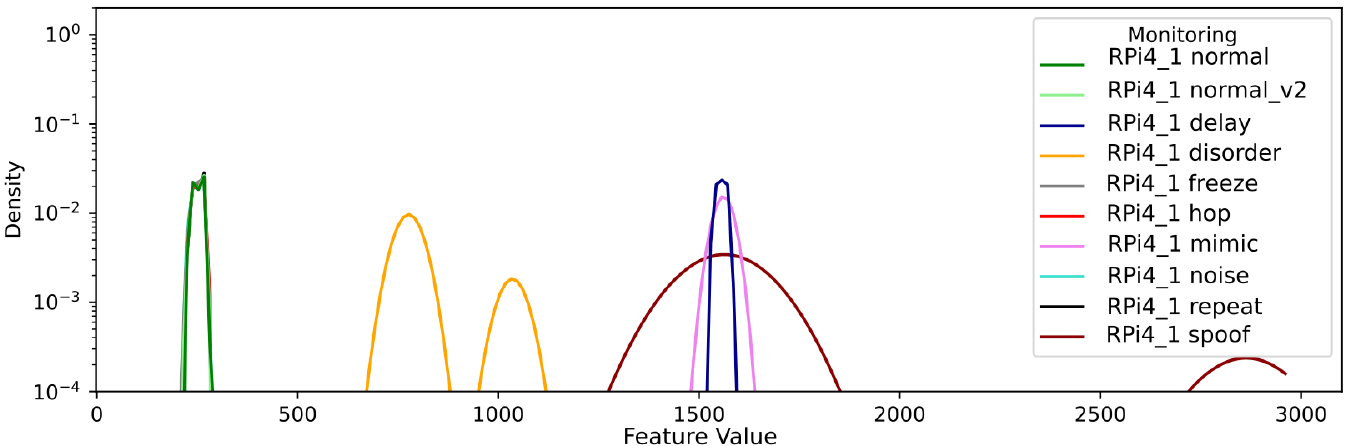}
    \caption{\textit{writeback\_mark\_inode\_dirty} Event of all Behaviors on RPi4\_1}
    \label{fig:attacks-writeback}
\end{figure}

To analyze the differences between normal and under-attack behaviors per device the distributions of each event have been studied. As a representative example, \figurename~\ref{fig:attacks-randomness} shows for RPi4\_1 and the \textit{urandon\_read} event how some attacks (hop, noise, and spoof) offer a higher number of random reads due to the generation of random noise. Another example can be seen in \figurename~\ref{fig:attacks-writeback}, where the \textit{writeback\_mark\_inode\_dirty} event of an RPi4\_1 is differently affected by the copy and swap operations of some SSDF attacks (being disorder the attack with the lowest impact on the event values). 


From the previous data exploration, it can be concluded that attacks do generally not impact the same features equally across different device types. Therefore, generalization across attacks and device types is challenging, and ML/DL/FL usage seems adequate for finding the events and values separating normal and SSDF attacks. In terms of data distribution, the exploration shows that the IID data is present in devices of the same device type and between the RPi4 families. However, devices from RPi3 and RPi4 families present non-IID data. Furthermore, the independence of the data samples allows that the data of a single device can be used to simulate additional participants of the same device type, which is critical for federated scenarios. Finally, external factors like network outages could potentially affect the data distributions. However, no significant systematic influence of external factors has been identified during the data exploration.

\section{Federated Attack Detection}
\label{sec:federated}

This section evaluates the performance of different federated models when detecting SSDF attacks and compares it with centralized ML/DL approaches where data privacy is not preserved. 

For that, two approaches have been considered. The first one detects anomalies using an unsupervised Autoencoder, while the second utilizes a supervised multilayer perceptron (MLP) to classify normal and under-attack behaviors. The pipeline and methodology followed to train and evaluate the federated models are also detailed in this section. Finally, four scenarios with different federation compositions (in terms of number and type of participants, behaviors, and detection tasks) are created to evaluate the performance of the previous FL models and compare it with centralized ML/DL approaches.  

\subsection{Federated ML Pipeline}
\label{sec:federated-ml-pipeline}

The federated setting needs adaptations from the typical ML pipeline to handle distributed data and models. In particular, the scaling phase and the threshold selection have to be adapted to allow a global model to aggregate the knowledge of involved participants. Furthermore, a central coordinator needs to run the federated learning pipeline iteratively. The following subsections describe the necessary steps.

\subsubsection{Dataset Splitting and Feature Preprocessing}

Each participant of the federation creates the following datasets: one for training, one for validation and optimization of hyperparameters, and another for testing the model performance. Those datasets are sampled from the respective dataset to avoid overlapping between sets. Next, outlier filtering is performed on the training and validation sets using the \textit{z-score}. \textit{Z-score} is computed using the mean $\mu$ and the standard deviation $\sigma$ according to the formula $\frac{x - \mu}{\sigma}$. Data points that have an absolute \textit{z-score} $\geq3$ in any feature are excluded as they could impair the model performance. Besides, features with correlation of 1 in the datasets are filtered.

\subsubsection{Federated Feature Scaling}

Feature scaling in a federated setup does not require communication efforts, as a global scaling for all participants must be put in place. Min-max scaling is employed using the formula $\frac{x - min}{max-min}$, but the minimum and maximum values are determined on the data of all the participants. Therefore, action from a central entity is required to coordinate the scaling process, asking the minimum and maximum of each feature to each participant and then returning the global minimum and maximum for scaling. A drawback of this approach is a certain loss of privacy since every participant has to disclose the minimum and maximum value of each feature. This issue could be addressed using solutions such as homomorphic encryption, but it is out of the scope of this work.

\subsubsection{Model Setup, Training and Evaluation}

Throughout this work, both supervised and unsupervised models are evaluated. It requires different data and methods to train the models and make predictions. However, both models are trained on a 68-dimensional input, which corresponds to the number of relevant features after the preprocessing. \textit{Stochastic gradient descent (SGD)} is used as the optimization algorithm with a learning rate of 1e-3 and a momentum term of 0.9. 

In the \textit{anomaly detection scenarios}, an Autoencoder with a single hidden layer of size 32 is used. After the first linear layer, \textit{batch normalization} is applied and \textit{GELU} is used as an activation function on the hidden state. A second linear layer transforms the hidden state back to its original size, followed by a \textit{GELU} activation function that returns the reconstructed input. After the training phase, the anomaly threshold is determined based on the mean ($\mu$) and standard deviation ($\sigma$) of the reconstructed mean square error (MSE). The formula used to select the threshold is show in Equation \ref{eq:thr}.

\begin{equation} \label{eq:thr}
    threshold = \mu + 3 \cdot \sigma
\end{equation}

The prediction then corresponds to determining the MSE on reconstructing a given behavioral vector. If the MSE of the recreated input is greater than the threshold, it is considered an anomaly and, therefore, behavior under attack. Otherwise, it is considered normal behavior.

In the \textit{binary classification scenarios}, an MLP is used. A linear layer produces a hidden state of size 256. Subsequently, batch normalization and the \textit{GELU} activation function are applied to this hidden state. A second linear layer then returns a single output neuron. A \textit{Binary Cross Entropy} Loss function with logits (\textit{BCEwithLogitsLoss}) is used during training, which applies the \textit{sigmoid} activation function and minimizes the logarithmic difference of the output to the encoded label (0 for normal behavior and 1 for attack behavior). Early stopping is applied when there is no loss decrease greater than \textit{1e-4} on the validation set.

For the federated training, \textit{FederatedAveraging (FedAvg)} algorithm is used. Algorithm \ref{alg:fedavg} describes the training loop in the clients and the server. Generally, the federation is trained for 15 aggregation rounds with five local epochs per participant if not stated otherwise. It is important to note that the models are relatively small and thus can also be trained on resource-constrained hardware. Further, early stopping is implemented per participant. 

\begin{algorithm}[ht!]
\begin{algorithmic}
 \STATE \textbf{Server executes:}
 \item  \hskip1em initialize $w_0$ 
 
 \item  \hskip1em \textbf{for} each round \textit{t} = 1,2... \textbf{do}
 \item  \hskip2em $m\leftarrow max(C \cdot K,1)$
 \item  \hskip2em $S_t \leftarrow$ (random set of $m$ clients)
 \item  \hskip2em \textbf{for} each client $k \in S_t$ \textbf{in parallel do}
 
 \item  \hskip3em $w^k_{t+1} \leftarrow $ ClientUpdate($k,w_t$)
 
 \item  \hskip2em $w_{t+1} \leftarrow \sum_{k=1}^{K} \frac{n_k}{n} w_{t+1}^k$ //Aggregation\\

 \item  \textbf{ClientUpdate(}$k,w$\textbf{):} //\textit{Run on client} $k$ 

 \item  \hskip1em $B \leftarrow$ (split $P_k$ into batches of size $B$) 
 
 \item  \hskip1em \textbf{for} each local epoch \textit{i} from 1 to \textit{W} \textbf{do}
 \item  \hskip2em \textbf{for} batch \textit{b} $\in B$ \textbf{do}

  \item \hskip3em $w \leftarrow w - \eta \bigtriangledown l(w;b)$ //Local update 

 \item \hskip1em return $w$ to server

 \end{algorithmic}
 \caption{\texttt{FederatedAveraging}. The \textit{K} clients are indexed by \textit{k}; \textit{B} is the local minibatch size, \textit{E} is the number of local epochs, and $\eta$ is the learning rate; $w$ are the model weights; $P_k$ is the local dataset of client $k$. \cite{mcmahan2017communication}}
 \label{alg:fedavg}
\end{algorithm}

\subsubsection{Federated Threshold Selection}
\label{subsec:fed-threshold-selection}

For anomaly detection, each participant sends its locally computed threshold to the central coordinator, which determines a global threshold. Depending on the federation composition, the thresholds per participant can vary heavily due to the non-IID data across different device types. It has to be taken into account when choosing the federated threshold. While a simple mean may perform reasonably in a setting with the same device type participants, it may perform poorly in a federation with different device types. Taking the maximum of the thresholds, on the other hand, creates a vulnerability to an overstating participant to impair the performance of the global model. Hence, a reasonable compromise has to be found. This compromise is built on the mean $\mu$ and standard deviation $\sigma$ of the list of thresholds that the participants send to the coordinator. Only thresholds that have an absolute \textit{z-score} that is $<=$ 1.5 are considered, choosing the maximum of those filtered values as the global threshold.

\subsection{Federated Scenarios and Detection Performance}

This section creates four federated scenarios where heterogeneous FL models are trained and evaluated following the previous pipeline. In addition, the detection performance of these models is compared to the one obtained by centralized approaches where data privacy is not preserved. For the sake of fair comparisons, both the federated and central models use the same algorithms, training and testing data, and hyperparameters. Finally, to show model performance and since the test sets for each behavior are separated, the accuracy of the model within each behavior test set is used.

The scenarios consider the devices and behaviors modeled by the dataset explained in Section~\ref{sec:dataset} to create the federations. To decide the number of sensors participating in each scenario, each participant must have enough data to achieve meaningful convergence in its local training loop. Therefore, the scenarios explained in this section restrict the number of participants per device type to a maximum of 4. Below, each scenario details the exact number and type of sensors used in its federation as well as the behaviors considered for training and testing.

\subsubsection{\textbf{Scenario 1: Federated Anomaly Detection with Balanced Device type}}

This scenario focuses on federated anomaly detection to detect zero-day attacks when there is a balanced federation of different sensors types (RPi3, RPi4 2GB, and RPi4 4GB). In particular, four participants of each sensor type are generated to set a total of 12. Between the 12 participants of the federation a privacy-preserving Autonencoder is trained following the pipeline previously explained. Each participant uses 1500 samples of its normal behavior for training and 150 different normal samples for the threshold selection task. Once the federated Autoencoder is trained, 75 samples per behavior (normal, normal\_v2, and eight SSDF attacks) of each participant are evaluated. 

\tablename~\ref{tab:scenario1} reports the accuracy achieved by the federated Autoencoder model per device type and behavior. The parentheses denote the difference with the accuracy of a central model (not protecting data privacy) that concatenates all train sets, and uses the same algorithm and hyperparameters. A positive difference means that the federation outperforms the simple central approach, whereas a negative difference is the opposite. Finally, it is important to note that RPi4\_2 is excluded from training and only used for testing.

\begin{table}[ht]
    \caption{Accuracy of Scenario 1 Autoencoder Model and Difference with a Centralized Approach (in Parentheses)}
    \centering
    \begin{adjustbox}{width=\columnwidth}
    \begin{tabular}{l|l|l|l|l}
     Behavior   & RPi3 (diff.)        &  RPi4\_1 (diff.)     & RPi4\_2 (diff.)      & RPi4\_3 (diff.)       \\
    \hline
    \hline
     normal     & 96.0\% (2.7\%)  & 100\% (1.3\%) & 100\% (2.7\%) & 99.3\% (0.7\%)  \\
     normal\_v2  & 96.0\% (3.3\%)  & 96.7\% (2.7\%)  & 99.3\% (0.7\%)  & 98.7\% (6.7\%)  \\
     \hline
     delay      & 100\% (0\%) & 100\% (0\%) & 100\% (0\%) & 100\% (0\%) \\
     disorder   & 100\% (0\%) & 100\% (0\%) & 100\% (0\%) & 100\% (0\%) \\
     freeze     & 0.7\% (-8.7\%)  & 4.00\% (-0.7\%)  & 2.0\% (0\%)   & 0\% (-1.3\%)  \\
     hop        & 100\% (0\%) & 100\% (0\%) & 100\% (0\%) & 100\% (0\%) \\
     mimic      & 100\% (0\%) & 100\% (0\%) & 100\% (0\%) & 100\% (0\%) \\
     noise      & 100\% (0\%) & 100\% (0\%) & 100\% (0\%) & 100\% (0\%) \\
     repeat     & 6.0\% (-4.0\%)  & 3.3\% (-0.7\%)  & 2.7\% (-2.0\%)  & 2.7\% (-2.0\%)  \\
     spoof      & 100\% (0\%) & 100\% (0\%) & 100\% (0\%) & 100\% (0\%) \\
    \hline
    \end{tabular}
    \end{adjustbox}
    \label{tab:scenario1}
\end{table}

As can be seen in \tablename~\ref{tab:scenario1}, both models (federated and centralized) perform almost identically, a good signal for the federated Autoencoder. More in detail, looking at the detection of anomalies produced by SSDF attacks, both models cannot detect freeze and repeat, but the rest of attacks are classified correctly ($\geq$96\%). An important aspect is that the accuracy on the second normal behavior (normal\_v2) is also high (96.00-99.33\%) despite not being used during training. 

\subsubsection{\textbf{Scenario 2: Federated Anomaly Detection with New Device type}}

This scenario evaluates whether a federated model can also be useful for a new device type joining the federation and detecting zero-day SSDF attacks. Thus, in this scenario, the federated anomaly detection model is trained with a total of eight participants belonging to two device types, and subsequently, the model is evaluated on behavior samples (normal and under-attack) of the new third device type. As indicated in \tablename~\ref{tab:models_sc2}, this is done for the three possible combinations of device types, generating three federated Autoencoder. In this context, following the previously defined pipeline each Autoencoder is trained. For each combination, participants provide 1500 samples of normal behavior for training, and 150 for the threshold selection. After that, 75 samples per behavior (two normal and eight attacks) of the third device type are used for testing.

\begin{table}[ht]
    \caption{Federated Models Used in Scenario 2 and 4}
    \centering
        \begin{adjustbox}{width=\columnwidth}
    \begin{tabular}{l|l|l}
     Model ID & Training Devices & Testing Devices \\
     \hline
     \hline
     Autoencoder/MLP 1  & RPi3 \& RPi4\_1  & RPi4\_3  \\
     Autoencoder/MLP 2  & RPi3 \& RPi4\_3  & RPi4\_1  \& RPi4\_2 \\
     Autoencoder/MLP 3  & RPi4\_1 \& RPi4\_3  & RPi3 \\
    \hline
\end{tabular}
    \end{adjustbox}
    \label{tab:models_sc2}
\end{table}

\tablename~\ref{tab:scenario2} shows the accuracy of the three federated Autoencoders and the difference with the centralized one. As an example, the first column displays the accuracy of Autoencoder 3 (see \tablename~\ref{tab:models_sc2}), trained with four participants of RPi4\_1 and four of RPi4\_3, and evaluated with the RPi3 samples. As in the previous scenario, RPi4\_2 is excluded from training and only used during testing.

\begin{table}[ht]
    \caption{Accuracy of Scenario 2 Autoencoder Models and Difference with a Centralized Approach (in Parentheses)}
    \centering
        \begin{adjustbox}{width=\columnwidth}
    \begin{tabular}{l|l|l|l|l}
     Behavior   & RPi3 (diff.)        &  RPi4\_1 (diff.)     & RPi4\_2 (diff.)      & RPi4\_3 (diff.)       \\
     \hline
     \hline
     normal     & 0\% (0\%)   & 98.0\% (4.0\%)  & 97.33\% (4.0\%)  & 99.3\% (9.3\%)  \\
     normal\_v2  & 0\% (0\%)   & 98.00\% (2.7\%)  & 99.3\% (4.0\%)  & 96.0\% (2.7\%)  \\
     \hline
     delay      & 100\% (0\%) & 100\% (0\%) & 100\% (0\%) & 100\% (0\%) \\
     disorder   & 100\% (0\%) & 100\% (0\%) & 100\% (0\%) & 100\% (0\%) \\
     freeze     & 100\% (0\%) & 4.7\% (-2.7\%)  & 0.7\% (-6.0\%)  & 0.7\% (-4.7\%)  \\
     hop        & 100\% (0\%) & 100\% (0\%) & 100\% (0\%) & 100\% (0\%) \\
     mimic      & 100\% (0\%) & 100\% (0\%) & 100\% (0\%) & 100\% (0\%) \\
     noise      & 100\% (0\%) & 100\% (0\%) & 100\% (0\%) & 100\% (0\%) \\
     repeat     & 100\% (0\%) & 2.00\% (-5.3\%)  & 4.00\% (-6.00\%)  & 1.3\% (-2.7\%)  \\
     spoof      & 100\% (0\%) & 100\% (0\%) & 100\% (0\%) & 100\% (0\%) \\
    \hline
\end{tabular}
    \end{adjustbox}
    \label{tab:scenario2}
\end{table}

As can be seen in \tablename~\ref{tab:scenario2}, knowledge transfer to unseen device types is possible if there are similarities in the hardware configuration. Since the behaviors of RPi3 and RPi4s are quite different (non-IID data), the knowledge transfer to RPi3 is not possible and all samples are classified as abnormal. In contrast, the performance on unseen RPi4s with different RAM is generally high, again with the exceptions of freeze and repeat behavior, which are not even detected when including the respective attack in the federation. Comparing the federated model to the centralized approach, there are no major differences, performing the federated model slightly better when detecting normal behavior.

\subsubsection{\textbf{Scenario 3: Federated Binary Classification with Balanced Device Type}}

It analyzes the capabilities of a federated binary classifier to transfer knowledge of known SSDF attacks between the federation. In particular, this scenario creates a federation of four participants per device type (12 in total) with the same behavioral data (normal and under-attack) per device type. More in detail, one participant per device type holds only normal data while the other three hold normal and delay, normal and freeze, and normal and noise, respectively. Each participant holds 250 samples of each selected behavior in its training set, 25 of each selected behavior in its validation set, and 75 of each existing behavior (two normal and eight attacks) in the test set. With this configuration and following the previous pipeline, a federated MLP is trained and evaluated. \tablename~\ref{tab:scenario3} shows the detection accuracy of the federated MLP model and the difference with the centralized approach. As usual, RPi4\_2 is only used during testing.

\begin{table}[ht]
    \caption{Accuracy of Scenario 3 MLP Model and Difference with a Centralized Approach (in Parentheses)}
    \centering
    \begin{adjustbox}{width=\columnwidth}
    \begin{tabular}{l|l|l|l|l}
     Behavior   & RPi3 (diff.)        &  RPi4\_1 (diff.)     & RPi4\_2 (diff.)      & RPi4\_3 (diff.)       \\
    \hline
    \hline
     normal     & 100\% (0\%)  & 100\% (0\%) & 100\% (0\%) & 100\% (0\%)  \\
     normal\_v2  & 100\% (4.0\%)  & 100\% (0\%) & 100\% (0\%) & 100\% (0\%)  \\
     \hline
     delay      & 100\% (0\%)  & 100\% (2.7\%) & 100\% (2.7\%) & 100\% (0\%)  \\
     disorder   & 93.33\% (49.3\%)  & 96.00\% (17.3\%) & 98.7\% (13.3\%) & 97.3\% (22.7\%)  \\
     freeze     & 0\% (-6.7\%) & 6.67\% (-4.67\%) & 5.33\% (-3.33\%) & 5.33\% (-4.67\%)  \\
     hop        & 100\% (0\%)  & 100\% (0\%) & 100\% (0\%) & 100\% (0\%)  \\
     mimic      & 100\% (0\%)  & 100\% (1.3\%) & 100\% (5.3\%) & 100\% (0\%)  \\
     noise      & 100\% (1.33\%)  & 100\% (0\%) & 100\% (0\%) & 100\% (0\%)  \\
     repeat     & 2.7\% (-2.7\%)  & 4.0\% (-2.6\%) & 5.3\% (-2.3\%) & 13.33\% (-0.5\%) \\
     spoof      & 100\% (0\%)  & 100\% (0\%) & 100\% (0\%) & 100\% (0\%)  \\
     \hline
\end{tabular}
    \end{adjustbox}
    \label{tab:scenario3}
\end{table}

As can be seen in \tablename~\ref{tab:scenario3}, the federated MLP transfers the attack knowledge quite well. It even improves the accuracy of a centrally trained model for the disorder attack. For behaviors other than disorder, no difference $>6\%$ can be observed between the federated and centralized approaches.

\subsubsection{\textbf{Scenario 4: Federated Binary Classification with New Device type}}

The last scenario is a combination of Scenarios 2 and 3. It evaluates the capabilities of a federated binary classifier to transfer attack knowledge from a federation to a new device type affected by attacks modeled in the federation. In particular, the scenario considers the same three federations of eight participants as Scenario 2 (see \tablename~\ref{tab:models_sc2}). In addition, as in Scenario 3, each participant holds 250 and 25 samples of selected behaviors for training and validation, respectively. Finally, the participant of the third device type holds 75 samples of each behavior (two normal and eight attacks) for testing. Following the pipeline previously explained, a federated MLP model per federation (3 in total) is trained. \tablename~\ref{tab:scenario4} shows the accuracy of the three federated MLP, and their differences with the centralized versions (using the same algorithms, data, and hyperparameters).

\begin{table}[ht]
    \caption{Accuracy of Scenario 4 MLP Models and Difference with a Centralized Approach (in Parentheses)}
    \centering
    \begin{adjustbox}{width=\columnwidth}
    \begin{tabular}{l|l|l|l|l}
     Behavior   & RPi3 (diff.)        &  RPi4\_1 (diff.)     & RPi4\_2 (diff.)      & RPi4\_3 (diff.)       \\
    \hline
    \hline
     normal     & 100\% (100\%) & 100\% (1.3\%) & 100\% (0\%) & 100\% (4.00\%)  \\
     normal\_v2  & 100\% (100\%) & 100\% (0\%) & 100\% (0\%) & 100\% (0\%)  \\
    \hline
     delay      & 0\% (-100\%)  & 100\% (0\%) & 100\% (0\%) & 100\% (0\%)  \\
     disorder   & 0\% (-100\%)  & 97.3\% (0\%)  & 100\% (1.3\%) & 88.0\% (-12.0\%) \\
     freeze     & 0\% (-2.3\%)  & 8.0\% (-3.0\%) & 6.67\% (-2.1\%) & 2.7\% (-6.4\%)  \\
     hop        & 1.33\% (-98.7\%)   & 100\% (0\%) & 100\% (0\%) & 98.7\% (1.3\%)   \\
     mimic      & 0\% (-100\%)  & 100\% (0\%) & 100\% (0\%) & 100\% (0\%)  \\
     noise      & 0\% (-100\%)  & 100\% (0\%) & 100\% (0\%) & 100\% (0\%)  \\
     repeat     & 1.3\% (-4.7\%)   & 4.0\% (-2.3\%) & 9.3\% (-2.0\%) & 5.3\% (-4.7\%)  \\
     spoof      & 0\% (-100\%)  & 100\% (0\%) & 98.7\% (-1.3\%) & 100\% (0\%)  \\
    \hline
    \end{tabular}
    \end{adjustbox}
    \label{tab:scenario4}
\end{table}

The results of \tablename~\ref{tab:scenario4} are very similar to those of Scenario 2. The transfer between RPi4s works well for most behaviors, while the transfer between RPi4 to RPi3 does not work at all (normal behavior of RPi3 is predicted as attack). Hence, there is no real advantage of either approach for the RPi3. Further, the model does not detect the behaviors freeze and repeat as attack in all RPi4. For these behaviors, the centralized model slightly outperforms the federated model. Further, the federated model shows better performance (+12\%) for the disorder behavior on a RPi4 with 4GB of RAM.

\section{Robustness Against Adversarial Attacks}
\label{sec:attacks}

This section evaluates the robustness of the FL models created in Scenario 1 and 3 (defined in Section~\ref{sec:federated}) when they are affected by malicious participants executing adversarial attacks. In particular, an increasing number of adversaries execute data and model poisoning attacks over a federated anomaly detector (Scenario 1) and a binary classifier (Scenario 3) equipped with different aggregation functions. 

In terms of adversarial attacks, the following are evaluated: \textit{(i)} behavior injections, as a variant of data poisoning, \textit{(ii)} model canceling, as a model poisoning attack, and \textit{(iii)} random weight upload, another model poisoning attack. More in detail, behavior injection uses malicious data to train models. In the case of anomaly detection, attack data is used as if it were normal, while for classification, the labels of normal and attack data are flipped during training. Model canceling tries to bring all global model parameters to zero. For that, it uploads the parameters of the last known global model multiplied by a factor $\alpha$ that is determined according to the formula based on the number of participants $K$ and the number of adversaries $f$: $  K - f + \alpha \cdot f = 0$. Finally, random weight upload sends random weights to the aggregation server. This work generates the weights from a normal distribution with a mean of zero and a standard deviation of three.

As secure aggregation functions, additionally to \textit{FedAvg}, the models consider \textit{(i) trimmed mean}, which excludes the highest and the lowest entries to calculate the mean of the local models weights, \textit{(ii) trimmed mean\_2} excluding the two highest and lowest entries for the averaging, and \textit{(iii) coordinate-wise median}, which uses the median of every weight instead of the average. Random weight upload uploads a random weight vector to the aggregation server. For the corresponding experiments random weight adversaries have been chosen to select their random weights using a normal distribution with a mean of 0 and a standard deviation of 3. Finally, to measure the impact of the attacks when evaluating normal and under-attack behaviors, all behavioral data of each participant are concatenated, and the F1-score metric is 
calculated as $F1-score=\frac{TP}{TP+\frac{1}{2}(FP+FN)}$ (TP:True Positive, TN:True Negative, FP: False Positive, FN: False Negative)

\subsection{Robustness of Scenario 1}

The previous adversarial attacks and secure aggregation functions are considered to measure the robustness of the federated Autoencoder detecting anomalies in Scenario 1.





\subsubsection{Attack Behavior Injection}

\begin{figure*}[t]
    \centering
    \includegraphics[width=1.7\columnwidth]{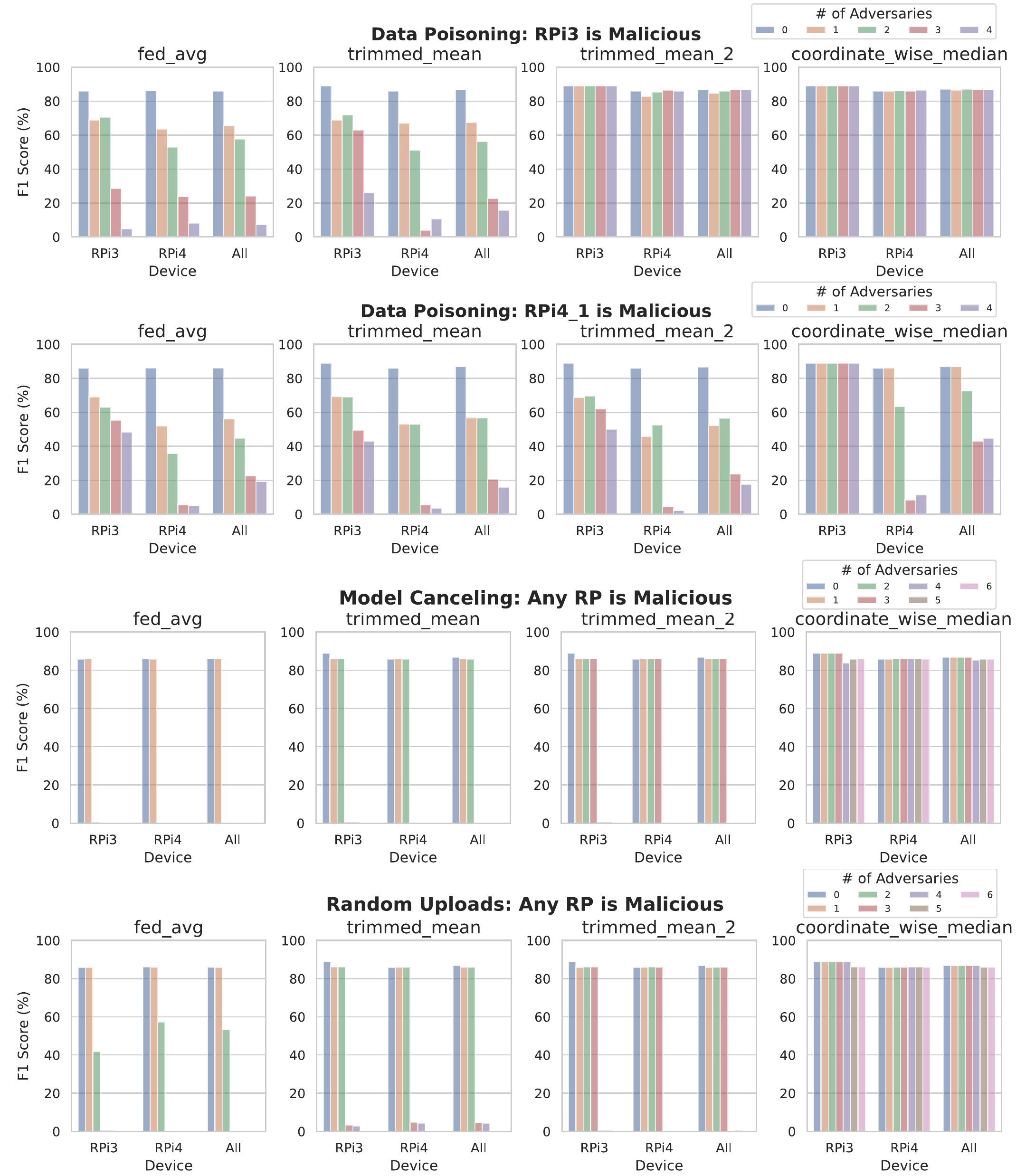}
    \caption{Impact of Different Adversarial Configurations on the Anomaly Detection Approach}
    \label{fig:ad_attacks}
\end{figure*}

In the federation of 12 participants (four per device type), from zero to four participants per device type (0\% to 33\% of the federation) are turned into data poisoning adversaries. This adversary setup is repeated three times, one per device type. Adversaries use attack samples (instead of normal samples) to train the federated Autoencoder. Each adversary injects different attack behaviors into the training process. In particular, the first adversary uses spoof behavior to train, the second mimic, the third delay, and the fourth disorder. These attacks are selected according to their median MSE in the corresponding federation without adversaries, choosing the ones more dissimilar from the normal behavior. Freeze and repeat behaviors are not injected due to their similarity to normal behavior. The first row of \figurename~\ref{fig:ad_attacks} shows for each device type, the F1-score of the federated Autoencoder according to the implemented aggregation function, and the number of adversaries belonging to the RPi3. The second row shows the same but when the adversaries belong to the RPi4 2GB family. Due to space constraints and similarities with the RPi4 2GB family, it is not shown when attackers belong to the RPi4 4GB type.

As can be seen in \figurename~\ref{fig:ad_attacks}, for federated averaging, even one adversary (8\% of the federation) decreases the F1-score of each device below 70\%. Four adversaries (33\%) destroy the model performance. Furthermore, the injecting device type matters and attacks performed by RPi4\_1 have a more significant impact on all different hardware configurations. Comparing the aggregation functions, \textit{coordinate-wise median} performs best in general, achieving an F1-score above 60\% for all test sets up to two adversaries. Especially in the case of adversarial RPi3, the aggregation function achieves excellent robustness with F1-scores above 80\% for up to 4 adversaries.

\subsubsection{Model Canceling and Threshold Attack}

In contrast to the previous attack, the impact of model canceling does not depend on the device type executing it. For this attack, the federation remains as in Scenario 1, but with up to 6 adversaries affecting the model robustness. It means that the number of participants varies from 12 (no adversaries) to 18 (with six malicious actors, 33\%). It is important to note that the model canceling attack is combined with an overstatement of the threshold. In other words, besides selecting model canceling weights, adversaries choose a threshold randomly from the uniform distribution in the range $[10^6, 10^9]$.

The third row of \figurename~\ref{fig:ad_attacks} reports the F1-score of each attack configuration per aggregation function. The \textit{FedAvg} aggregation is only capable of defending against one adversary. Most importantly, the threshold overstatement can destroy the model performance once one manipulated threshold is not filtered. In this scenario, the \textit{coordinate-wise median} provides a very robust defense since the federation maintains very good performance even with six adversaries. In conclusion, while the mean is shifted heavily towards the attackers for the \textit{FedAvg} and \textit{trimmed mean} aggregations, the median can be more stable against largely different adversarial model weights. However, in the case of $\geq50\%$ adversarial percentage, the median would also lose effectiveness.

\subsubsection{Random Uploads}

The fourth row of \figurename~\ref{fig:ad_attacks} shows the impact of the adversaries per aggregation function. Here, similar results as in the model canceling attack can be observed. \textit{Coordinate-wise median} performs best, followed by \textit{trimmed mean\_2} and the basic \textit{trimmed mean}. Nonetheless, in this attack where adversaries produce random weights, it is not as obvious as for model canceling how the aggregation function can filter the exact weight values. Random weights can be in a completely honest range for some layers or hidden units, but they can also be extreme values for others. It depends on which distribution the random values are sampled and whether they are extreme values compared to the honest weights. However, random weights have no significant impact on the median.

In conclusion, this scenario has shown how robust aggregation methods improve the model resilience against adversaries. In all attacks, \textit{coordinate-wise median} is the aggregation method offering the best robustness. It maintains the model performance almost unaltered in 3 of 4 adversarial attacks, only decreasing (still better than the other aggregation methods) when RPi4\_1 performs a data poisoning attack.

\subsection{Robustness of Scenario 3}

This section measures the robustness of the federated MLP classifying normal and under-attack behaviors in Scenario 3.


\subsubsection{Label Flipping}

For this attack, the federation of 12 participants (four per device type and three device types) contains from zero to four adversaries of a given device type. This adversary setup is repeated three times, once per device type. In each configuration, adversaries flip the labels of their local data. In particular, the first adversary flips the labels of normal behavior, the second flips normal and delay, the third normal and freeze, and the fourth normal and noise.

\begin{figure*}[t]
    \centering
    \includegraphics[width=1.7\columnwidth]{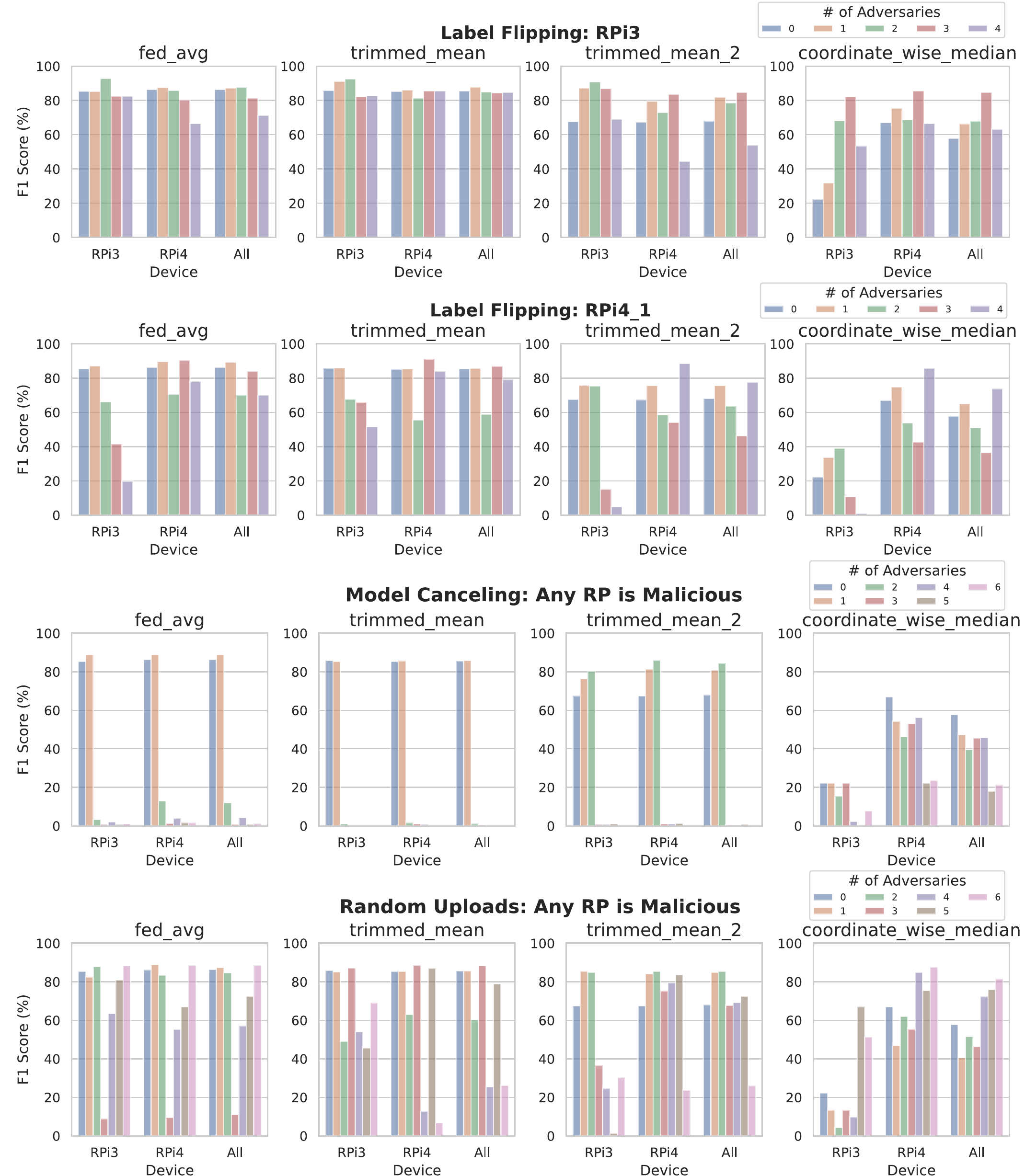}
    \caption{Impact of Different Adversarial Configurations on the Binary Classification Approach}
    \label{fig:binary_attacks}
\end{figure*}

The two first rows of \figurename~\ref{fig:binary_attacks} follow the same structure as \figurename~\ref{fig:ad_attacks}, but for label flipping attacks this time. As can be seen, for \textit{FedAvg} and regardless of the device type acting maliciously (but especially for RPi3 acting as an adversary), the attack does not have a big impact on the average model performance. This is because certain attack characteristics are already available in the federation. For example, if there are two adversarial RPi4\_1, the normal and freeze behavior labels are flipped for this specific device type only. However, the knowledge about these behaviors is fully present for the other two device types. It explains the much higher F1-score than in other attack scenarios. Apart from that, the \textit{trimmed mean} function is the one providing more robustness for all devices. In contrast, \textit{coordinate-wise median} shows a different pattern where the model performs poorly, especially for RPi3. It can be appreciated how performance even improves with the presence of some adversaries. However, once there are too many label flipping adversaries, it seems to become a lottery which participants weights are chosen for the update.

\subsubsection{Model Canceling}

This attack considers the same 12 participants of Scenario 3 and adds from zero to six adversaries (0-33\% of the federation). In contrast to the anomaly detection experiment, the threshold cannot be attacked in this case. The third row of \figurename~\ref{fig:binary_attacks} reports the results for zero to six model canceling adversaries regardless of the device type acting maliciously. As can be seen, while the performance for \textit{trimmed mean} excluding one extreme value is very similar to \textit{FedAvg} aggregation, the exclusion of two extreme values (trimmed mean\_2) helps to protect from one more adversary. Still, the performance drops below 20\% for three or more adversaries. \textit{Coordinate-wise median} performs better than the other aggregation functions with four or more adversaries but does not present a viable solution either. It might be explained by the fact that filtering out RPi3 good weights by the median as this device type represents a minority in the federation. 

\subsubsection{Random Uploads}

It also considers Scenario 3, where different adversaries (from zero to six) execute random weight model uploads. The fourth row of \figurename~\ref{fig:binary_attacks} reports the F1-score of the federated MLP for different aggregation functions and adversaries. As can be seen, the \textit{trimmed mean\_2} function provides the most robust results in general (and especially for RPi4). Adversaries generating random weights do not necessarily always produce wrong weights for the overall model. Moreover, it looks like random adversarial weights cancel out each other, which explains the instability. Indeed, the global model becomes more random when more adversaries are introduced. Therefore, federated averaging is highly unstable. The second \textit{trimmed mean} variant provides a better defense as more adversaries can be filtered, being especially in favor of the RPi4. Finally, \textit{coordinate-wise median} does not perform well for low numbers of adversaries but provides an effective countermeasure for four or more malicious participants.

In conclusion, this scenario has shown that robust aggregation methods are not as effective as in the first scenario. Here, there is not a clear aggregation method better than the others. \textit{Trimmed mean} is the one offering best results under label flipping attacks, while \textit{trimmed mean\_2} and \textit{coordinate-wise median} have the best results for model canceling and random uploads attacks, but still with a significant performance loss compared to no-attack situations.

\section{Summary, Conclusions, and Future Work}
\label{sec:conclusions}

This work evaluates the robustness of fingerprinting FL models equipped with anti-adversarial mechanisms and able to detect cyberattacks affecting resource-constrained devices. To achieve that goal, this work first creates and makes public a FL-oriented dataset based on fingerprints of Raspberry Pis utilized as spectrum sensors of ElectroSense. The dataset contains samples from eight different SSDF attacks and from two versions of normal behavior for a total of four physical sensors. After that, four federated scenarios based on anomaly detection and binary classification are created to evaluate and compare the detection performance of privacy-preserving FL models and DL models where data privacy is neglected. The main results of each scenario demonstrate that FL achieves a detection performance that can compete with centralized DL approaches without significant limitations. Finally, this work analyzes the impact of different amounts of malicious participants executing data and model poisoning attacks against FL models equipped with different aggregation mechanisms (\textit{FederatedAveraging}, \textit{trimmed mean}, and \textit{coordinate-wise median}). The experiments conducted show that both data poisoning and model poisoning severely affect the federated performance for binary classification as well as anomaly detection. However, \textit{Trimmed mean} and \textit{coordinate-wise median} can help if adversaries are present up to a certain percentage, but they cannot guarantee robustness, and their applicability depends on the specific scenario considered.

As the main conclusion of this study, FL legitimates itself by achieving competitive performance in scenarios where privacy plays an important role, meaning that a centralized setup is not possible. Due to its simplicity in training, the capability to detect zero-day attacks, and possible robustness improvements, anomaly detection appears to be the best solution for the federated detection of data integrity attacks in a crowdsensing platform. Regarding the robustness of FL models and the impact of anti-adversarial mechanisms, the evaluated aggregation functions provide different results depending on the scenario and attack. Generally, the \textit{FedAvg} is particularly vulnerable to extreme values as they distort the selected global weight average entirely. The \textit{trimmed mean} aggregation may be able to filter extreme values up to some extent, but if there are sufficiently many adversaries, not all weight uploads can be excluded. Further, there is the concern of excluding honest participants weights despite adversaries only. Thus, the best number of updates to exclude from the averaging is very difficult to determine in \textit{trimmed mean}. The \textit{coordinate-wise median aggregation} is most robust against extreme values but can lead to unstable results in heterogeneous federations.

As future work, there is still room for further research about robust aggregation mechanisms. In the case of anomaly detection, domain-specific aggregation functions could be used to filter adversaries more effectively by leveraging further knowledge about common distributions and fingerprint patterns of normal behavior. For instance, it could be utilized that the threshold of an honest federation participant should be in a certain range for a given device type. Lastly, a larger dataset could greatly enhance the exploration of the FL use case. This could allow to test more extensively how heterogeneity and non-IID data influence federated model performance.

\section*{Acknowledgment}

This work has been partially supported by \textit{(a)} the Swiss Federal Office for Defense Procurement (armasuisse) with the CyberTracer and RESERVE projects (CYD-C-2020003) and \textit{(b)} the University of Zürich UZH.

\bibliographystyle{apalike}  
\bibliography{references}

\begin{IEEEbiography}[{\includegraphics[width=1in,clip]{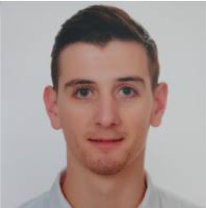}}]{Pedro M. Sánchez Sánchez} received the MSc degree in computer science from the University of Murcia, Spain. He is currently pursuing his PhD in computer science at University of Murcia. His research interests are focused on continuous authentication, networks, 5G, cybersecurity and the application of machine learning and deep learning to the previous fields.
\end{IEEEbiography}

\begin{IEEEbiography}[{\includegraphics[width=1in,clip]{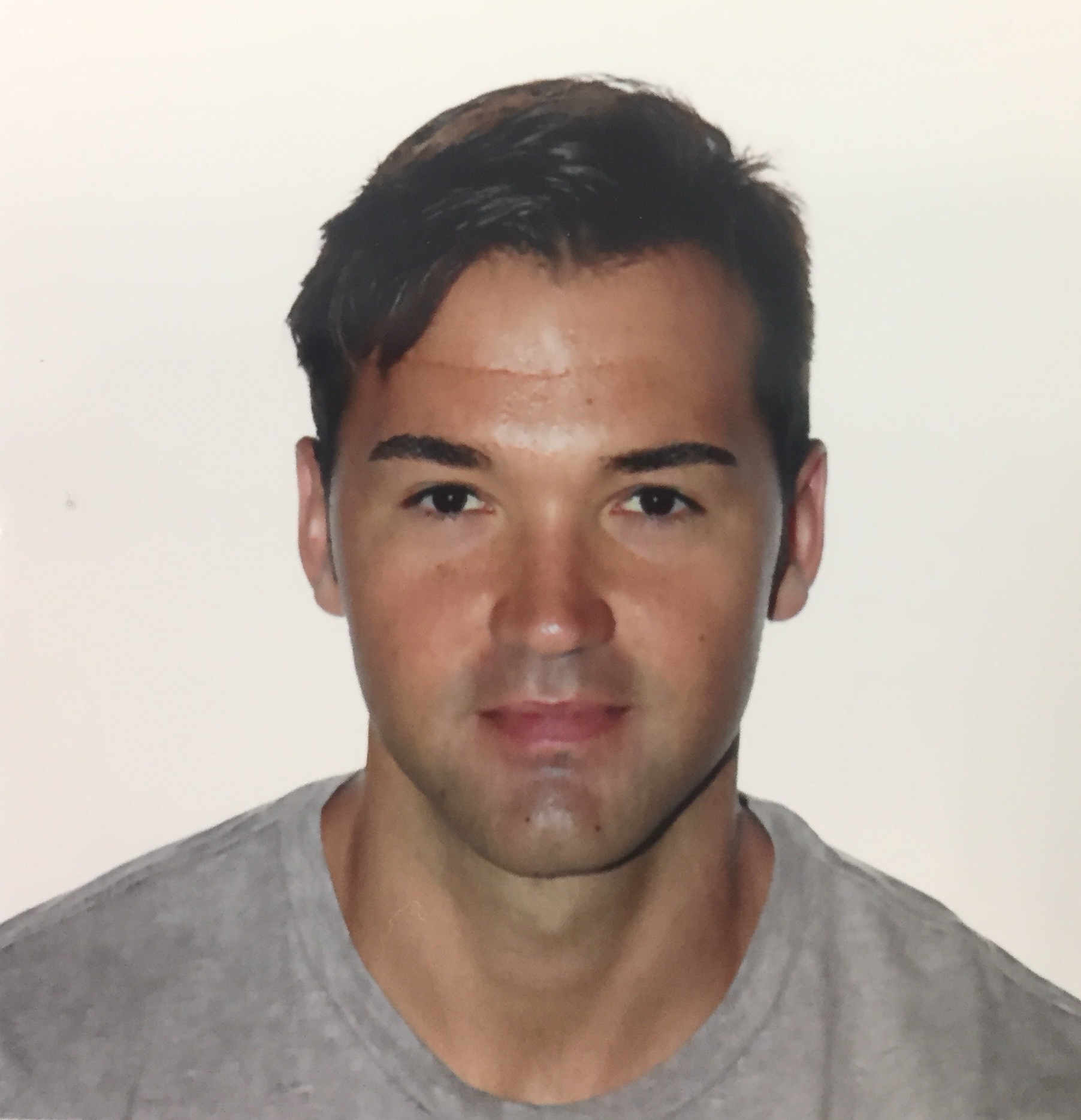}}]{Alberto Huertas Celdrán} received the MSc and PhD degrees in Computer Science from the University of Murcia, Spain. He is currently a postdoctoral fellow at the Communication Systems Group CSG, Department of Informatics IfI at the University of Zurich UZH. His scientific interests include IoT, BCI, cybersecurity, data privacy, continuous authentication, semantic technology, and computer networks.
\end{IEEEbiography}
 
\begin{IEEEbiography}[{\includegraphics[width=1in,height=1.25in,clip]{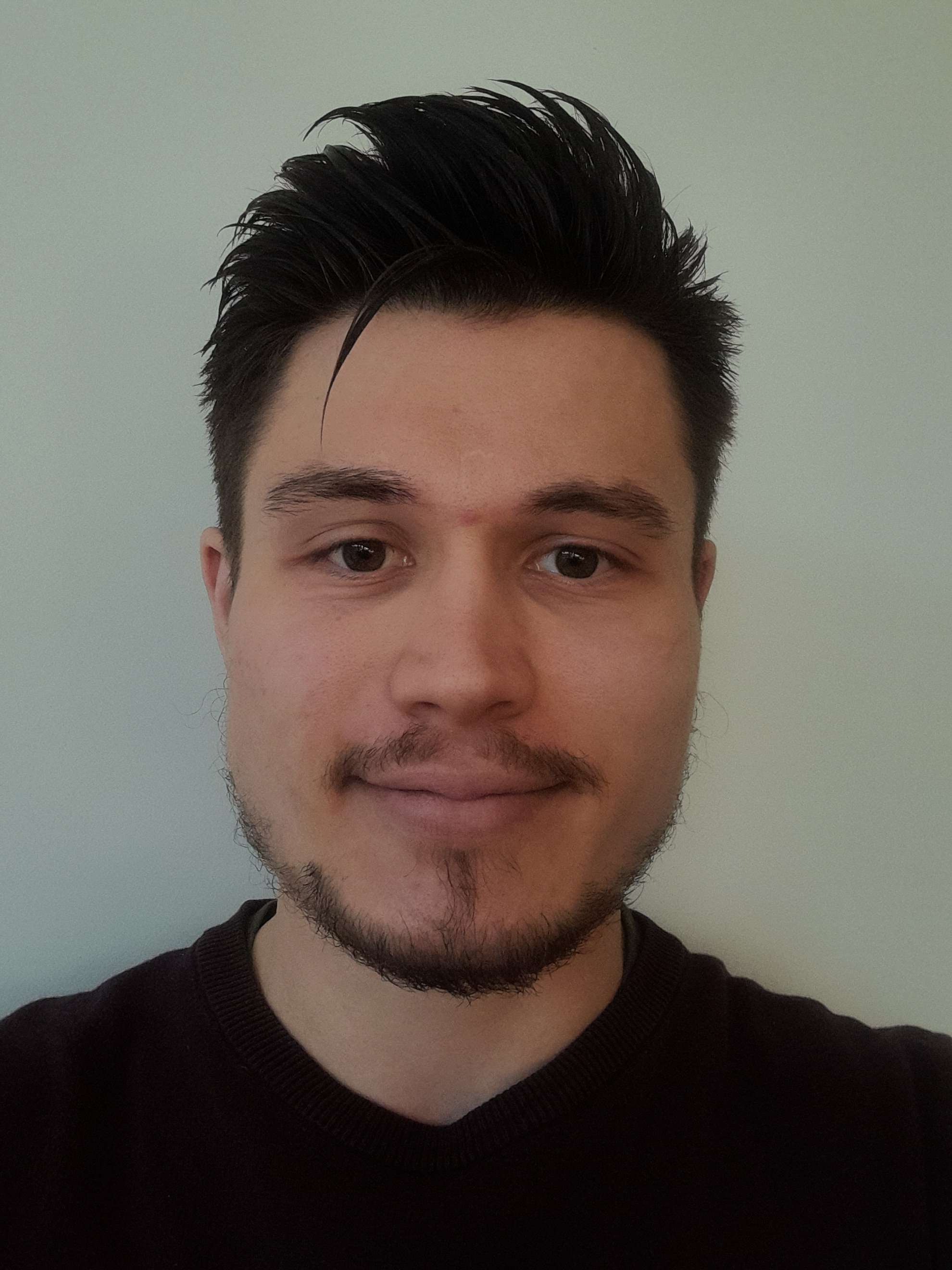}}]{Timo Schenk}  is currently completing his MSc degree in Computer Science at the University of Zurich UZH, Switzerland. While having received his BSc degree from the Department of Informatics IFI UZH, he has also gained experience across multiple positions in the software engineering industry. He is passionate about cybersecurity and artificial intelligence and has a particular scientific interest in application areas where these two fields intersect.
\end{IEEEbiography}

\begin{IEEEbiography}[{\includegraphics[width=1in,clip]{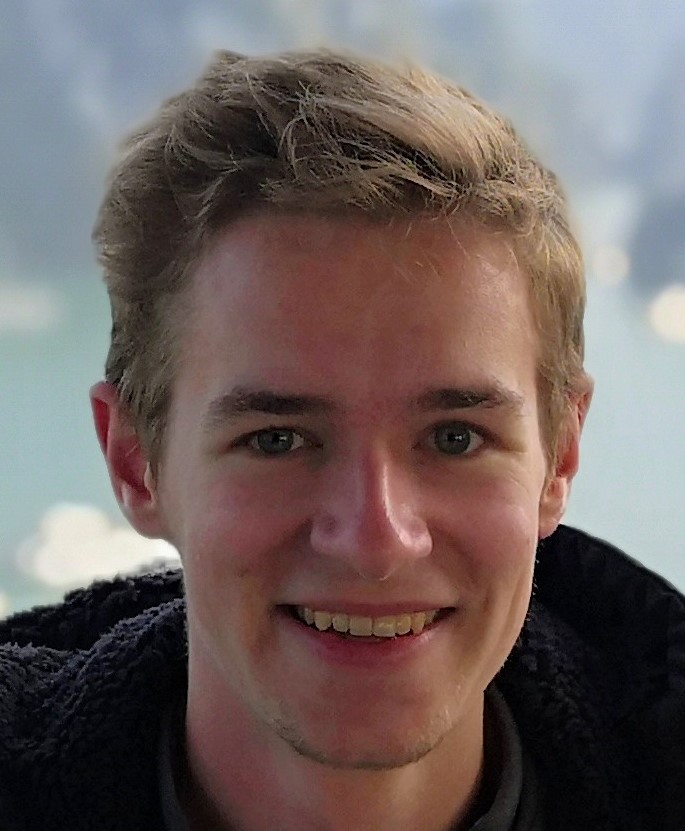}}]{Adrian Lars Benjamin Iten} received his BSc in Computer Science from the University of Zurich, Switzerland evaluating compact Machine Learning (ML) models for Natural Language Processing (NLP). He is currently pursuing his MSc in Software Systems at the University of Zurich and has previously worked as a software engineer in the industry for five years. His scientific interests include the applications of ML for cyber security and NLP.
\end{IEEEbiography}

\begin{IEEEbiography}[{\includegraphics[width=1in,clip]{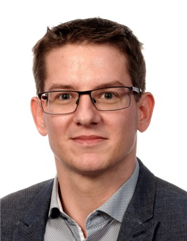}}]{Gérôme Bovet} is the head of data science for the Swiss Department of Defense, where he leads a research team and a portfolio of about 30 projects in the context of Cyber-Defence and intelligence. His work focuses on Machine and Deep Learning approaches, with an emphasis on anomaly detection, adversarial and collaborative learning applied to data gathered by IoT sensors. He received his Ph.D. in networks and computer systems from Telecom ParisTech, France, in 2015, and an Executive MBA from the University of Fribourg, Switzerland in 2021.
\end{IEEEbiography}

\begin{IEEEbiography}[{\includegraphics[width=1in,clip]{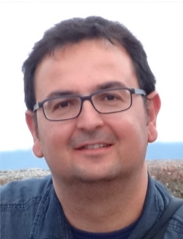}}]{Gregorio Martinez Pérez} is Full Professor in the Department of Information and Communications Engineering of the University of Murcia, Spain. His scientific activity is mainly devoted to cybersecurity and networking. He is working on different national (14 in the last decade) and European IST research projects (11 in the last decade) related to these topics, being Principal Investigator in most of them. He has published 160+ papers in national and international conference proceedings, magazines and journals.
\end{IEEEbiography}

\begin{IEEEbiography}[{\includegraphics[width=1in,clip]{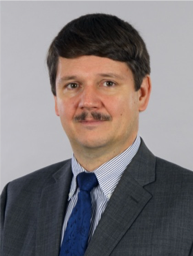}}]{Burkhard Stiller} received the Informatik-Diplom (MSc) degree in Computer Science and the Dr. rer.- nat. (PhD) degree from the University of Karlsruhe, Germany, in 1990 and 1994, respectively. Since 2004 he chairs the Communication Systems Group CSG, Department of Informatics IfI, University of Zürich UZH, Switzerland as a Full Professor. His main research interests are published in well over 300 research papers and include systems with a fully decentralized control (blockchains, clouds, peer-to-peer), network and service management, Internet-of-Things, and telecommunication economics.
\end{IEEEbiography}

\end{document}